\documentclass[1p, final]{elsarticle} 

\usepackage{lineno,hyperref}
\usepackage{graphicx}
\usepackage{dcolumn}
\usepackage{bm}
\usepackage{color}
\usepackage{graphicx}
\usepackage{subcaption}
\usepackage{tabularx}
\usepackage{array}
\usepackage{multirow}
\usepackage{mathtools}
\usepackage[export]{adjustbox}   
\usepackage{mwe}    
\usepackage{tikz}
\usetikzlibrary{shapes, backgrounds}
\usepackage{booktabs}
\usepackage{float}
\usepackage{nomencl}
\usepackage{upgreek}

\modulolinenumbers[5]

\journal{Mechanics of Materials}

\begin{document}
\graphicspath{{./figures/}}
\begin{frontmatter}

\title{Stochastic mechanical modeling of metallic foams to determine onset of mesoscale behavior}

\author[mymainaddress]{Mujan N. Seif\corref{mycorrespondingauthor}}
\cortext[mycorrespondingauthor]{Corresponding author}
\ead{mujan.seif@uky.edu}
\author[secondaryaddress]{Jake Puppo}
\author[secondaryaddress]{Metodi Zlatinov}
\author[secondaryaddress]{Denver Schaffarzick}
\author[thirdaddress]{Alexandre Martin}
\author[mymainaddress]{Matthew J. Beck}

\address[mymainaddress]{Department of Chemical and Materials Engineering, College of Engineering, University of Kentucky, Lexington, KY, USA}
\address[secondaryaddress]{ERG Aerospace, Oakland, California, 94608}
\address[thirdaddress]{Department of Mechanical and Aerospace Engineering, College of Engineering, University of Kentucky, Lexington, KY, USA}

\begin{abstract}
Metallic foams are crucial to many emerging applications, among them shielding against hypervelocity impacts caused by micrometeoroids and orbital debris. 
The variability of properties at feature-scale and mesoscale lengths originating from the foam's inherently random microstructure makes predictive models of their properties challenging.
It also hinders the optimization of components fabricated with such foams, an especially serious problem for spacecraft design where the balance between cost and mass must also be balanced against the catastrophic results of component failure.
To address this problem, we compute the critical transition length between the feature-scale, where mechanical properties are determined by individual features, and the mesoscale, where behavior is determined by ensembles of features.
At the mesoscale, distributions of properties--with respect to both expectation value and standard variability---are consistent and predictable.
The Kentucky Random Structure Toolkit (KRaSTk) is applied to determine the transition from feature-scale to mesoscale for computational volumes representing metallic foams at a range of reduced densities.
The transition is found to occur when the side length of a cubic sample volume is $\sim$10$\times$ greater than the characteristic length.
Comparing KRaSTk-computed converged stiffness distributions with experimental measurements of a commercial metallic foam found excellent agreement for both expectation value and standard variability at all reduced densities. 
Lastly, we observe that the diameter of a representative MMOD strike is $\sim$30$\times$ shorter than the feature-scale to mesoscale transition for the foam at any reduced density. 
Therefore, features will determine response to hypervelocity impacts, rather than bulk (or even mesoscale) structure.
 

\end{abstract}


\end{frontmatter}

\linenumbers

\nolinenumbers
\section{Introduction}
\begin{figure}[bh]
    \centering
    \includegraphics[width=1\linewidth]{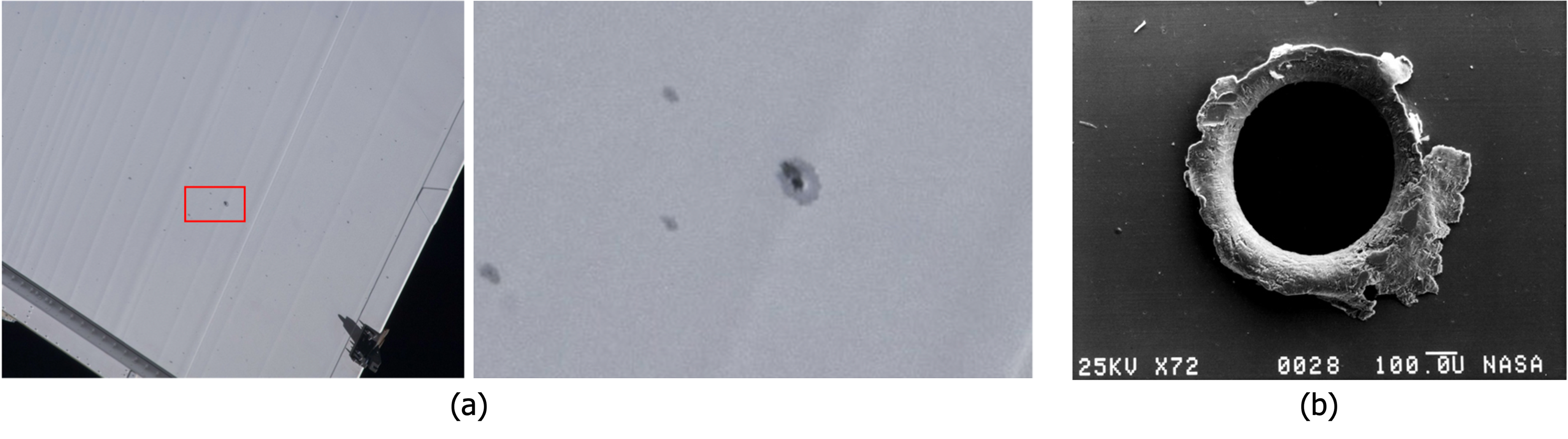}
    \caption{Examples of MMOD impacts: \textbf{(a)} Surface of an ISS radiator panel, reproduced from Ref.~\cite{Hyde:NA:2019} and \textbf{(b)} MMOD impact on satellite component retrieved during the STS-41C Solar Max repair mission, reproduced from Ref.~\cite{NASA:N:2013}.}
    \label{fig:mmodimpacts}
\end{figure}

Micrometeoroids and orbital debris (MMOD) poses a significant danger to both robotic and crewed spacecraft operating beyond Earth's atmosphere \cite{Christiansen:NA:2009,Klavzar:PE:2015,Buenrostro:JCM:2018,Ryan:NA:2010,Zhang:PE:2017,Kader:IJIE:2016}.
Despite sizes under 2 mm, even miniscule particles can have tremendous impact velocities (10-12 km/s or Mach $\sim$35, equivalently) that result in catastrophic damage \cite{Christiansen:NA:2009}. 
While spacecraft are generally maneuvered to avoid impacts with objects large enough to be detected and tracked, this is impossible with small particles, and collisions are inevitable (Fig.~\ref{fig:mmodimpacts}).
Therefore, crews, payloads, and instrumentation must be protected with some form of energy absorbing and damage mitigating shielding that must itself be aggressively optimized to minimize both mass and volume.

Open cell metallic foam sandwich panel structures have proven effective at mitigating damage from MMOD strikes \cite{Buenrostro:JCM:2018, Ryan:NA:2010,Zhang:PE:2017}. 
Commercially available metallic foams sandwiched between solid metallic facesheets are widely used for MMOD shielding on a range of spacecraft. 
The foam core layer is composed of open, node-ligament networks of randomly oriented polycrystalline metallic ligaments.
Ligament lengths and pore sizes vary over some distribution characteristic of the specific material, but typically have lengths on the order of millimeters.
Impacting particles penetrate the outer sacrificial facesheet of the sandwich structure and suffer repeated impacts with ligaments in the foam---deflecting, fragmenting, and vaporizing in the process.
Any residual impactor fragments arriving at the rear facesheet have sufficiently reduced kinetic energy that they can be stopped without the danger of perforation.

Operating under powerful imperatives to minimize weight and volume while robustly assuring mission success (within externally determined parameters), component and spacecraft designers require accurate knowledge of foam properties and responses to MMOD strikes.
This is complicated by the potentially small sizes of impacting MMODs and the intrinsically random nature of foam-based shielding materials, which exhibit different materials responses depending on the length-scale of the particle/foam interactions.
For impacts affecting volumes at size scales similar to that of individual ligaments and/or pores---the ``feature" length scale, on the order of 100s to 1000s of microns for many metallic foams---foam properties and responses to MMOD strikes are not characteristic of the bulk foam material, but rather of individual structural features (e.g., pores, ligaments, ligament junctions, etc.) within the structure.
At length scales much greater than those of individual pores and ligaments---that is, on the macroscale---uniform, singular bulk properties characteristic of the foam material can be observed and represent the homogenized, or ``effective'', response of all structural features (ligaments and pores).
Between these length scales---on the mesoscale---material properties become characteristic of the foam material, but are not singular or uniform, instead varying within characteristic distributions depending on the detailed arrangement of features in the affected volume.

In order to make appropriate design decisions---particularly ones that minimize mass and volume---spacecraft designers require quantitative knowledge of (i) expected bulk materials properties, (ii) distributions of local properties characteristic of the material at the mesoscale, and (iii) the physical lengths which bound the feature-scale, mesoscale, and macroscale.
In particular, it is critical to know the length scale above which behavior representative of the bulk material will be observed---that is, the transition length from the feature-scale to mesoscale---as this sets a hard minimum size for spacecraft components fabricated from the material.
This need for quantitative knowledge of materials properties and responses has led to high demand for accurate, physics-based computational models allowing rapid materials selection (among, e.g., foams with different ligament and pore sizes) and computational prototyping of component performance.
To meet these needs, models of complex and/or randomly structured materials must necessarily account for the intrinsic inhomogeneity of the bulk material and allow efficient sampling of the materials structure at a range of length scales.

The Kentucky Random Structures Toolkit (KRaSTk) was recently developed to address the need for accurate, physics-based, multiscale modeling of complex and randomly structured materials.
The KRaSTk approach has been described previously \cite{Seif:AM:2021} and focuses on leveraging a geometric seed description of a material with complex structure to generate and compute properties of many model representative volume elements (mRVEs). 
By computing properties of many mRVEs at various sizes relative to the key structural features of the seed description, this approach allows for determination of bulk effective properties, distributions of mesoscale properties, and the size at which the material's behavior will transition from feature-scale to mesoscale.
Here we apply KRaSTk to compute the properties of mRVEs designed to model metallic foams and validate computed results by comparing to independently obtained experimental results for a commercially available Al foam (Duocel, manufactured by ERG Aerospace) \cite{ERGAerospace:NA:2021}.
We find that a node-ligament seed geometry accurately predicts foam properties and extends experimentally available data by providing distributions of expected properties relevant at the mesoscale.
We further demonstrate that sample sizes of at least ten times typical microstructural features sizes are required to yield characteristic materials behavior (as opposed to behavior characteristic of individual ligaments and voids).

\section{Methods}
\subsection{mRVE Generation}
\begin{figure}
    \centering
    \includegraphics[width=1\linewidth]{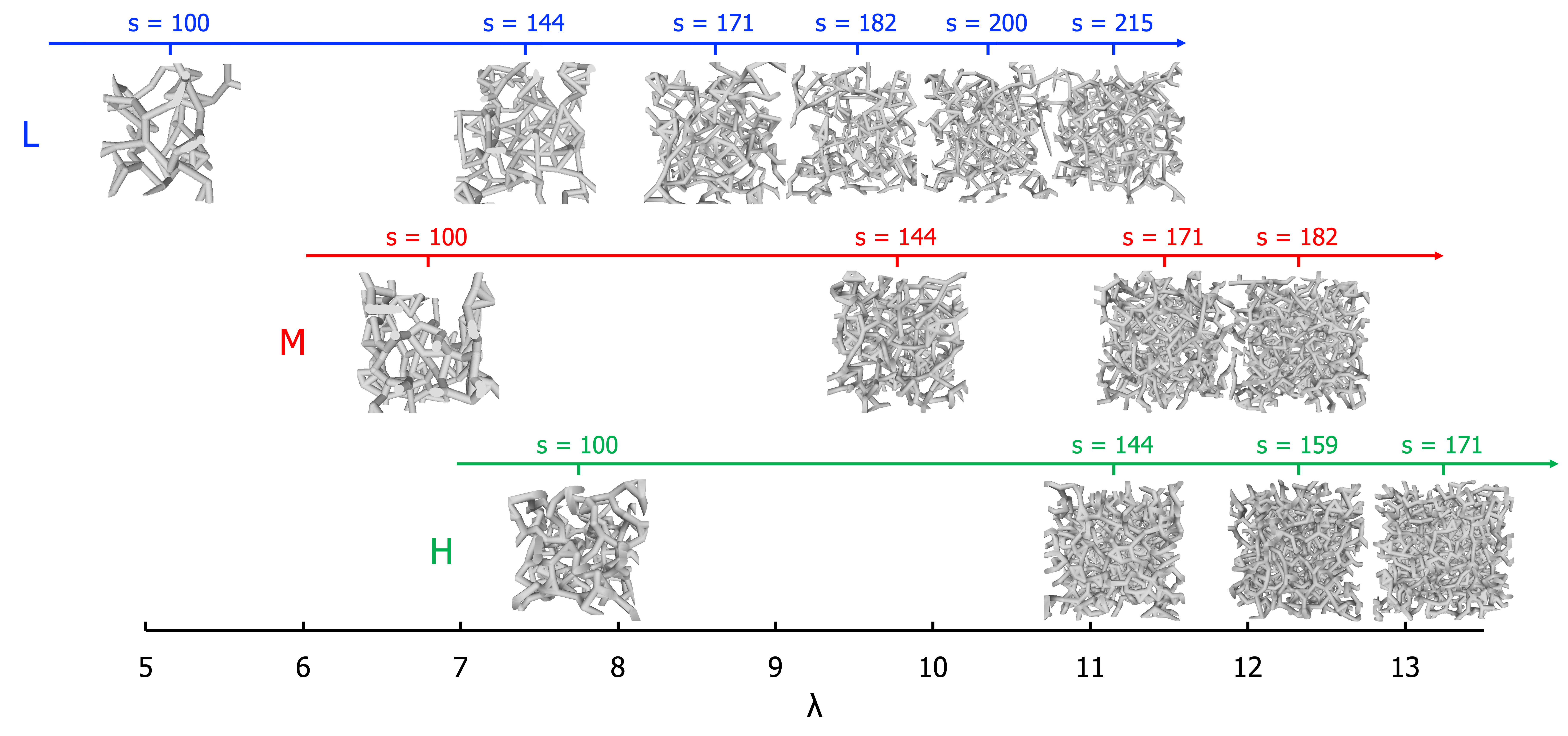}
    \caption{A ``map" of each set of structures utilized in this study. It is important to note that the structure shown for each set of conditions is just an example mRVE; in reality, 200 mRVEs were generated for each set.}
    \label{fig:mRVEsets}
\end{figure}

The Kentucky Random Structures Toolkit (KRaSTk) \cite{Beck:NA:NA,Seif:AM:2021} was used to generate sets of 200 model representative volume elements (mRVEs) to characterize metallic foams (Fig.~\ref{fig:mRVEsets}).
KRaSTk procedurally generates large numbers of stochastic model representative volume elements (mRVEs) based on a geometric seed description of the characteristic structural features defining the material in question.
Individual mRVEs are volumes that stochastically sample arrangements of structural features possible based on the seed geometry.
Sets of mRVEs generated with common parameters (e.g., sizes or lengths of characteristic features) can then be used to quantitatively predict sample size dependent properties of real materials whose structure is described by the seed geometry.

In connecting computed properties of mRVEs to properties of real materials, it is important to determine the length scales bounding the macroscale, mesoscale, and feature-scale.
For any particular material these critical transition lengths are, necessarily, relative to the size of characteristic features in the material.
To account for this, the parameter $\lambda$ is used to represent the relative size of an mRVE, where $\lambda$ is defined as the ratio of the length of an mRVE to characteristic feature lengths in the seed geometry of interest \cite{Seif:AM:2021}.
Leveraging this terminology, we refer specifically to the feature-scale to mesoscale transition length as $\lambda_{f \rightarrow m}$.

The prototype geometric seed used here to represent metallic foams consisted of randomly placed spherical nodes connected by conical frustum ligaments \cite{Seif:AM:2021}.
The spherical nodes have radii randomly distributed within a user-specified range and were separated by a user-specified minimum distance.
Conical frustum ligaments with circular cross-sections were attached to nodes, which determine the end radii of each ligament.
A stochastic network was constructed by creating ligaments connecting each spherical node to a user-specified minimum number of nearest neighbor nodes ($N_C$).
To represent a real volume removed from a larger, continuous sample, mRVEs with volume $V$ were generated by adding nodes and ligaments to a cube of volume $2V$. 
The cube-shaped final mRVE of volume $V$ was then cut from the center of the $2V$ domain, truncating ligaments and nodes at the mRVE faces as would be observed in a physical sample cut from a larger bulk.
The average length of ligaments in each mRVE ($\ell$) is taken to be the characteristic geometric feature length, and the cube root of the mRVE volume is taken to be the mRVE length (hence, for cubic mRVEs, this means that $\lambda=s/\ell$, for $s$, the mRVE side length).

For this node-ligament seed geometry, mRVEs can be generated with different ligament thicknesses (by using different node diameters), average ligament lengths (different volume densities of nodes and therefore greater separation between nearest neighbor nodes), and degree of constraint in the ligament network (different $N_C$)--factors all combining to result in a range of possible reduced densities.
It is not generally possible to independently vary all of these parameters, as they are related in non-trivial ways---i.e., conserving volume per ligament by varying average ligament length and ligament diameter in a reciprocal way does not conserve reduced density, as the number of ligaments (and nodes) in a given volume decreases (increases) as the average ligament length increases (decreases).
Highlighting this interdependence, if the number of spherical nodes in an mRVE with volume $V$ increases, the nodes are closer together, and ligament lengths decrease. 
For fixed ligament diameters, the ligament aspect ratio will decrease and reduced density increases. 
The same change in ligament aspect ratio and reduced density can be achieved by holding the node density constant (and therefore ligament lengths constant) while increasing ligament diameters.
This also highlights that lengths in mRVEs from the same seed geometry are relative---that is, mRVE size is described by $\lambda$, ligament sizes by aspect ratio, and mRVEs may be scaled arbitrarily by a linear factor.
In this study, mRVEs with three reduced densities (high, medium, and low)---each with two different network connectivities ($N_C=3$ or $4$)---were generated.
All mRVEs were constructed with ligament diameters of 6 units, and different reduced densities were achieved by varying the volume density of nodes (and therefore the average ligament length).
For mRVEs with $N_C=3$, structures with different $\lambda$ were constructed by varying the mRVE side length, $s$.
A total of fourteen sets of mRVEs with $N_C=3$ were built: six with low reduced densities ($\sim$4.2\%) and $\lambda = 5.1$, 7.4, 8.8, 9.4, 10.3, and 11.1; 
four with medium reduced density ($\sim$8.1\%) and $\lambda = 6.8$, 9.7, 11.6, and 12.3; and four with high reduced density ($\sim$12.0\%) and $\lambda = 7.7$, 11.1, 12.3, and 13.2.
Three sets with $N_C=4$, again distinguished as low, medium, and high reduced density, were generated with $\lambda = 10.3, 11.6$, and 11.1, respectively.
This yielded sets with reduced densities of 5.9\%, 10.7\%, and 15.1\%.
The average ligament lengths ($\ell$) for all mRVEs---regardless of $N_C$, which will not affect $\ell$---in the low, medium, and high reduced density sets were 19.4, 14.9, and 13.0, respectively.
Given that the constituent ligaments in all mRVEs were constructed with a fixed diameter, these average ligament lengths yielded aspect ratios (AR) of 3.2 (low density), 2.4 (medium density), and 2.2 (high density).
In the following, sets of mRVEs are labelled as X/$\lambda$, where ``X'' is H, M, or L for high, medium, and low reduced density, and the $\lambda$ value for the set is indicated.

\subsection{Computing mRVE Properties via FEM}
The orthorhombic stiffness tensor for each mRVE was computed using an FEM-based approach, extending a previously developed method \cite{Seif:AM:2021}.
To extract the full stiffness tensor, total elastic energies were computed for each mRVE in nine independent strain states established using combinations of displacement and frictionless support boundary conditions. 
The nine strain states included the three independent uniaxial compressions ($\epsilon_i=\bar{\epsilon}\neq0, \epsilon_j=\epsilon_k=0$), three biaxial compressions ($\epsilon_i=\epsilon_j=\bar{\epsilon}\neq0, \epsilon_k=0$), and three pure shear strains.
All applied strains had magnitudes of $\bar{\epsilon}=0.5$\%.
Computed total strain energies for the nine strain states yield a system of equations of the form:
\begin{align}
    U_{ij} &= \frac{1}{2}C_{ij}\epsilon_i\epsilon_j \label{eq:strainenergy}
\end{align}
allowing direct calculation of C$_{11}$, C$_{22}$, C$_{33}$, C$_{12}$, C$_{13}$, C$_{23}$, C$_{44}$, C$_{55}$, and C$_{66}$ in terms of the nine computed $U_{ij}$ values for each mRVE.
Stiffness constants were then converted to elastic moduli (E$_1$, E$_2$, E$_3$) Poisson's ratios ($\nu_{12}, \nu_{13}, \nu_{32}$), and shear moduli ($G_{23}, G_{13}, G_{12}$), as described in the Appendix and Ref.~\cite{Bower:NA:2009}.

Finite element calculations were conducted using the \texttt{FEniCSx} general PDE solver package \cite{Alnaes:ANS:2015,Logg:ATMS:2010}. 
A Krylov solver incorporating the generalized minimal residual method and a successive over-relaxation pre-conditioner was used.
The absolute tolerance was set to $1\times10^{-7}$ and the relative tolerance to $1\times10^{-5}$.
Displacement boundary conditions were used to apply strains appropriate to each of the nine considered strain states.
The material within each ligament was chosen to be isotropic, with bulk intrinsic mechanical properties of E = 69 GPa and $\nu$ = 0.33, similar to reported values for polycrystalline Al. 
Geometries were meshed utilizing \texttt{mshr}, the mesh generation component of \texttt{FEniCSx}.
\texttt{Mshr} builds simplicial \texttt{DOLFIN} meshes in 3D, utilizing \texttt{CGAL} and \texttt{Tetgen} as mesh generation backends.
For each structure, a tetrahedral mesh with a resolution of 150.0 was generated.
The resolution quantity is defined as the inverse of the cell size $h$, the longest edge in any mesh element.
Solid fractions were computed as the total mesh volume. 

\subsection{Determining materials properties from mRVE properties}
FEM calculations result in computed stiffness tensors which are distinct for each mRVE.
The properties of the material represented by the chosen seed geometry are characterized by the distribution of computed properties for a sufficiently large set of mRVEs--that is, computed distributions of properties for at least $N_{min}$ mRVEs represent a quantitative prediction of distribution of measured properties that would be observed for a volume of material with the same size (as measured in terms of $\lambda$).
A previous study using the same geometric seed considered here found $N_{min}$ to be 30 mRVEs \cite{Seif:AM:2021} and sets of $N=200$ mRVEs were used here to ensure robust computed distributions.
To characterize and aid analysis of distributions of elastic moduli for sets of mRVEs, results are represented as $\Gamma$-distributions---which were selected because they are positive definite (as elastic modulus must be) and converge to normal distributions for expectation values (mean values) far from zero.
$\Gamma$-distributions for each set of mRVEs are fit to computed elastic moduli using the \texttt{GammaDistribution} object in the Statistics and Machine Learning Toolbox in \texttt{MATLAB}, resulting in shape ($a$) and scale ($b$) parameters characterizing each mRVE set.
Expectation values for $\Gamma$-distributions of $E$ values are $\mu_E=a*b$, and standard deviations of $E$ values within mRVE sets are $\sigma_E=(a*b^2)^{1/2}$.

At the feature-scale (that is, for volumes with small $\lambda$ values), sets of $N>N_{min}$ mRVEs generated with the same structural parameters will not yield the same expected properties ($\mu_E$ values), despite being converged with respect to the number of sampled structures.
This is because mRVEs at this scale do not represent the material as a bulk, rather they represent arbitrary (but not representative) combinations of individual features within the material.
Above a critical $\lambda_{f\rightarrow m}$ value, $\mu_E$ for different mRVE sets (generated with the same structural parameters and having $N>N_{min}$) will converge, indicating a transition to the mesoscale, where mRVE sets do effectively characterize the material of interest.
At this size scale, mRVE sets will still have non-zero $\sigma_E$ values (standard deviations of computed elastic moduli), but these non-zero standard deviations in computed elastic moduli do not represent computational error, rather they characterize the variability of local properties intrinsic to the material itself at the length scale ($\lambda$ value) considered.
To highlight this we define the range $[\pm \sigma_E]$ to be the \emph{standard variability} of the material, and this range, which is a materials property, depends on $\lambda$.
As discussed above, for mRVEs at extremely large $\lambda$ values the standard variability (that is, the intrinsic variability in the observed properties of the material) will fall to zero, representing the transition to the macroscale---the size scale at which the material exhibits singular bulk properties.
While $[\pm \sigma_E]$ will fall to zero in the macroscopic limit, the decay with increasing $\lambda$ is expected to be extremely slow for the range of $\lambda$ values considered in this study--all of which are expected to fall within, or just below, the mesoscale.



\section{Results \& Discussion}
\subsection{Feature-scale to mesoscale transition}
\begin{figure}
    \centering
    \includegraphics[width=1\linewidth]{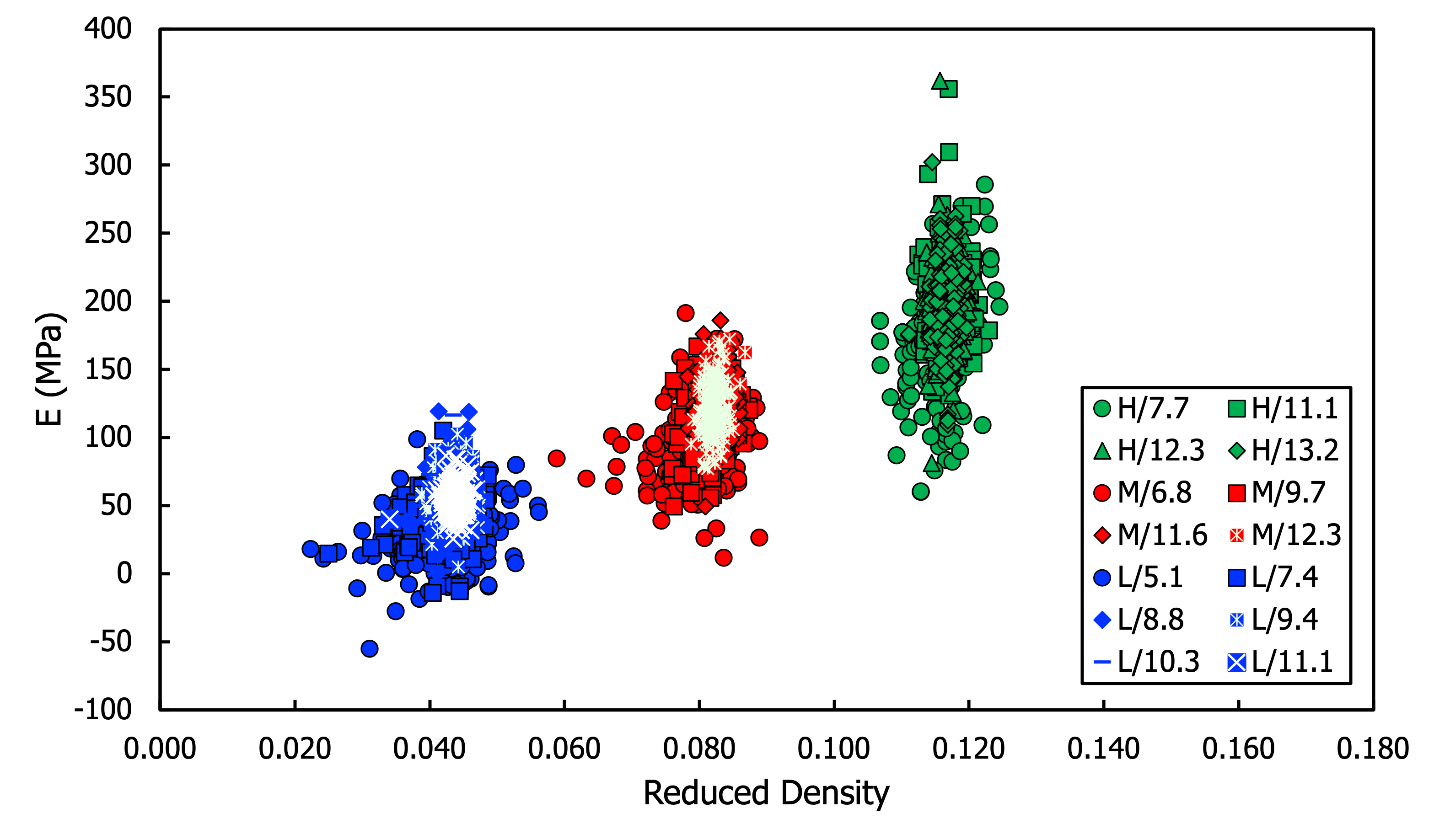}
    \caption{$E$ results for each mRVE set represented as a scatter plot.}
    \label{fig:fullscatter}
\end{figure}

\begin{figure}
    \centering
    \includegraphics[width=1.0\linewidth]{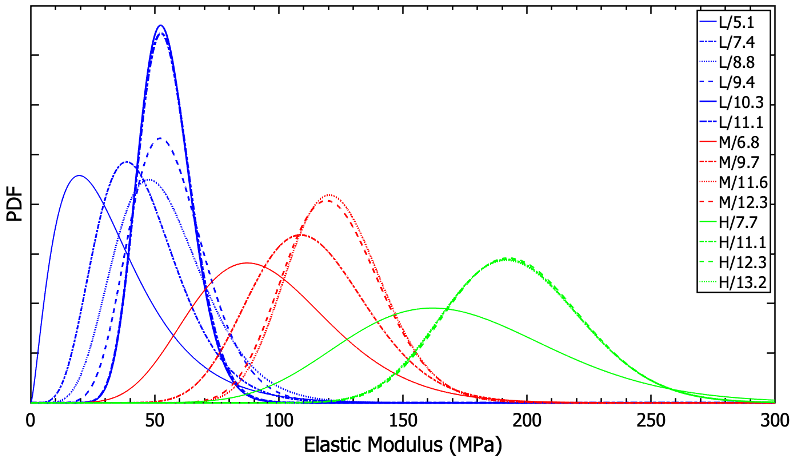}
    \caption{$E$ results for each mRVE set, plotted as gamma distributions. The gamma parameters and statistical quantities are found in Table~\ref{tab:allgammasdistparameters}.}
    \label{fig:alldists}
\end{figure}

The elastic moduli computed for each $N_C=3$ mRVE considered in this study are shown in Fig.~\ref{fig:fullscatter}.
Different color markers indicate $E$ values for mRVEs with high (H), medium (M), or low (L) densities, while different marker shapes indicate sets of mRVEs with different $\lambda$.
Fig.~\ref{fig:alldists} shows the same results as Fig.~\ref{fig:fullscatter}, though represented as individual $\Gamma$-distributions for each mRVE set. 
Careful examination of Fig.~\ref{fig:fullscatter} reveals that a small number of mRVEs, all within the lowest $\lambda$ low density mRVE set, yielded non-physical negative E values.
These negative stiffness values occur in instances where there are so few ligaments in a (small) mRVE that it is mechanically unstable.
Negative E values were omitted when computing and plotting the corresponding positive-definite $\Gamma$-distribution.

\begin{figure}
    \centering 
    \includegraphics[width=1\linewidth]{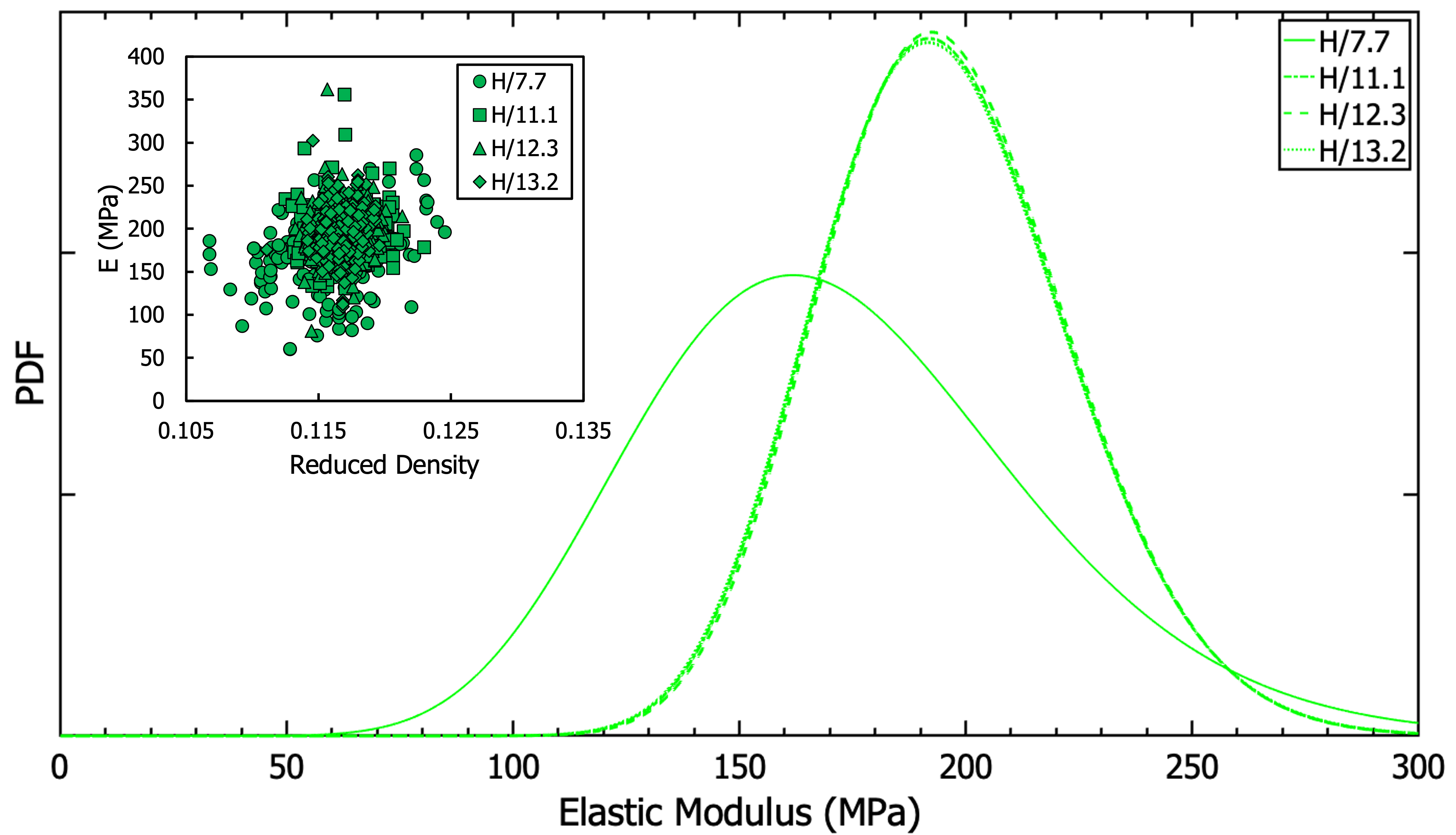}
    \caption{$E$ results for the high density sets, plotted as both scatter and gamma distributions. The gamma parameters and statistical quantities are found in Table~\ref{tab:allgammasdistparameters}.}
    \label{fig:highdensresults}
\end{figure}

Fig.~\ref{fig:highdensresults} shows data for high density mRVEs (sets H/7.7, H/11.1, H/12.3, and H/13.2) in the form of $\Gamma$-distributions (with the raw point clouds of computed values shown in the inset), which highlight the results' convergence with increasing $\lambda$ to an expected bulk effective stiffness value, $\mu_E$.
Beginning with H/11.1, both $\mu_E$ and $[\pm \sigma_E]$ are extremely consistent as the size ($s$, and therefore $\lambda$) of the mRVEs increases.
Table~\ref{tab:allgammasdistparameters} also highlights the change, with both $\mu_E$ and $[\pm \sigma_E]$ differing for $\lambda = 7.7$ but converged for all other cases\footnote{At much higher $\lambda$ values (much larger computational boxes compared to ligament lengths) $[\pm \sigma_E]$ will approach zero, while $\mu_E$ will remain fixed, representing the transition from mesoscale to bulk behavior. 
For the \emph{range} of $\lambda$ values considered here, the feature-scale to mesoscale transition, $[\pm \sigma_E]$ is expected to vary so slowly as to appear converged in the mesoscale.}.
This convergence indicates the transition from the feature-scale to the mesoscale and sets a range on the value of $\lambda_{f\rightarrow m}$ for high density structures: $7.7 < \lambda_{f\rightarrow m} \leq 11.1$.

\begin{table}[]
    \centering
    \caption{The gamma distribution parameters, $a$ and $b$, and corresponding physical quantities, $\mu_E$ and $\sigma_E$, of each mRVE set with $N_C=3$.}
    \begin{tabular}{ccccc}
        \toprule
        & \multicolumn{2}{c}{$\Gamma$ Parameters} & \multicolumn{2}{c}{Elastic Modulus} \\
        \cmidrule(lr){2-3}\cmidrule(lr){4-5}
        Set & $a$ & $b$ & $\mu_E$ & $\sigma_E$ \\
        \midrule
        L/5.1 & 2.41 & 13.87 & 33.36 & 21.51\\ 
        \midrule
        L/7.4 & 6.66 & 6.81 & 45.35 & 17.57\\ 
        \midrule
        L/8.8 & 8.35 & 6.48 & 54.04 & 18.71 \\ 
        \midrule
        L/9.4 & 13.31 & 4.24 & 56.41 & 15.47 \\ 
        \midrule
        L/10.3 & 26.03 & 2.09 & 54.40 & 10.66 \\ 
        \midrule
        L/11.1 & 25.15 & 2.17 & 54.68 & 10.90 \\ 
        \midrule
        M/6.8 & 10.65 & 9.04 & 96.32 & 29.51  \\ 
        \midrule
        M/9.7 & 22.39 & 5.08 & 113.83 & 24.06 \\
        \midrule
        M/11.6 & 41.00 & 3.01 & 123.37 & 19.27\\
        \midrule
        M/12.3 & 38.18 & 3.21 & 122.42 & 19.81\\ 
        \midrule
        H/7.7 & 16.16 & 10.68 & 172.58 & 42.94\\ 
        \midrule
        H/11.1 & 49.40 & 3.96 & 195.82 & 27.86\\ 
        \midrule
        H/12.3 & 50.59 & 3.88 & 196.29 & 27.60\\ 
        \midrule
        H/13.2 & 48.75 & 4.01 & 195.49 & 28.00\\
        \bottomrule
    \end{tabular}
    \label{tab:allgammasdistparameters}
\end{table}

The same procedure was followed to determine $\lambda_{f\rightarrow m}$ for the low and medium density mRVE sets with $N_C=3$.
In addition to the $\Gamma$-distributions plotted in Fig.~\ref{fig:alldists}, $\mu_E$ and $[\pm \sigma_E]$ values are reported in Table~\ref{tab:allgammasdistparameters}.
For medium density structures, $\mu_E$ and $[\pm \sigma_E]$ converge as $\lambda$ approaches 11.6, indicating that $9.7 < \lambda_{f\rightarrow m} \leq 11.6$ in this case.
The results for the low density mRVE sets differ in that there is a differentiation between the convergence of the expectation value, $\mu_E$, and standard variance ($[\pm \sigma_E]$) as $\lambda$ increases. 
Beginning with L/9.4 and persisting through L/10.3 and L/11.1, $\mu_E$ is around 55 MPa (see Table~\ref{tab:allgammasdistparameters}). 
However, it is not until L/10.3 that a consistent $[\pm \sigma_E]$ emerges.
This highlights that accurately resolving the variability intrinsic to a material with complex structure (at a particular $\lambda$) requires larger (relative to the feature sizes) volumes than resolving the expectation value of the property.

\begin{figure}
    \centering
    \includegraphics[width=1\linewidth]{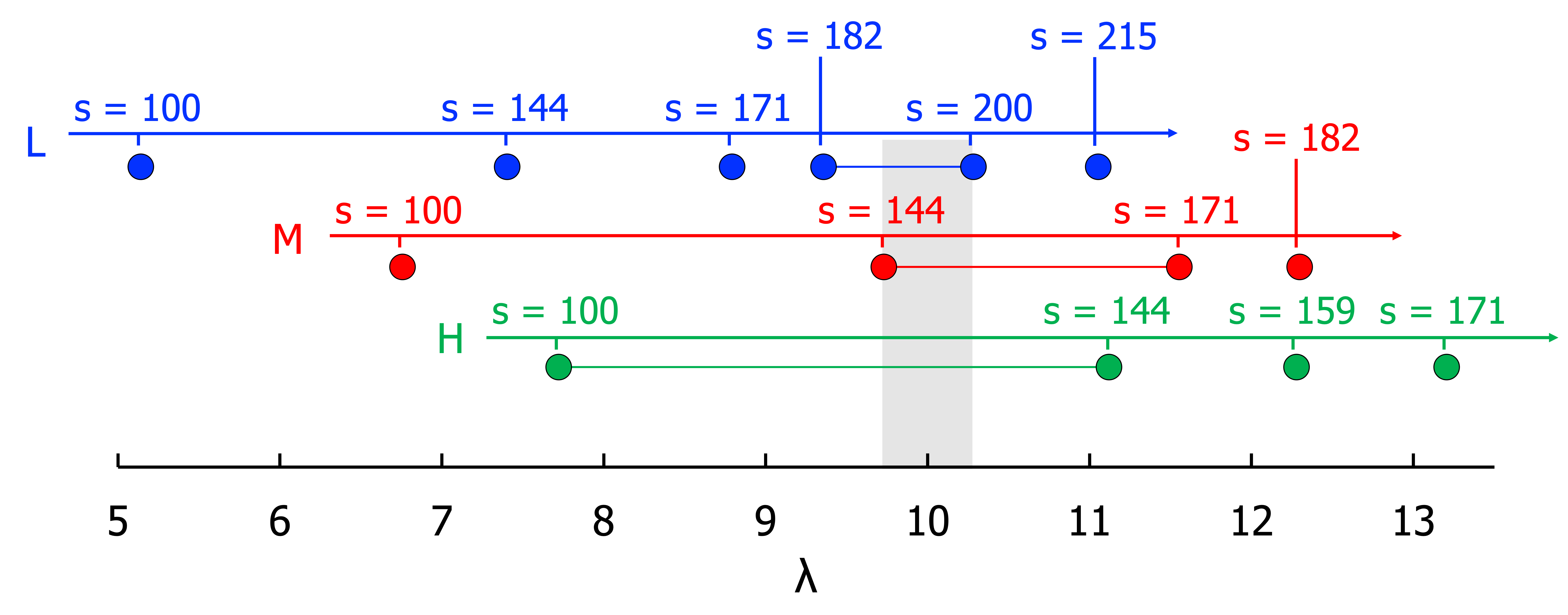}
    \caption{The quantity $\lambda$ is plotted for each mRVE set. Within each class of mRVEs, the range of $\lambda$ where $\mu$ and $\sigma$ converge below 2\% is highlighted with a colored line segment. The gray shaded region indicates the range of $\lambda$s where the converged windows of each class overlap. }
    \label{fig:alllambdanumberline}
\end{figure}
Fig.~\ref{fig:alllambdanumberline} indicates the $\lambda$ values for all sets and aggregates the ranges established for $\lambda_{f\rightarrow m}$.
It is immediately evident that computed ranges for $\lambda_{f\rightarrow m}$ overlap in the range $9.7 \leq \lambda_{f\rightarrow m} \leq 10.3$.
This demonstrates that the relative length scale at which foam materials behave as materials (as opposed to individual ligaments and pores---the feature-scale to mesoscale transition length) does not depend on reduced density.
It also shows that $\lambda_{f\rightarrow m}$ does not depend on on the average ligament aspect ratio, as the various high, medium, and low density mRVEs were constructed by varying the ligament length while holding the ligament diameter constant.
Combined, this indicates that the length scale at which a foam transitions to mesoscale behavior is universal for self-similar foams---that is, for foams described by the same seed geometry and network connectivity.

While the \emph{relative} length scale for the feature-scale to mesoscale transition is common to all structures with the same seed geometry, because self-similar foams with different average ligament lengths ($\ell$) can be fabricated, the \emph{physical sample size} implied by a universal $\lambda_{f\rightarrow m}$ varies from sample-to-sample.
The present calculations highlight this as the different reduced densities considered were achieved by varying the volume density of nodes in different mRVEs, necessarily resulting in different average ligament lengths in low, medium, and high reduced density mRVEs.
In fact, the side lengths required to achieve mesoscale behavior ($s_{f\rightarrow m}$ as analogous to $\lambda_{f\rightarrow m}$) in the low, medium, and high density mRVEs studied here varied by approximately 50\%:
\begin{align}
    \textrm{Low}: 188.2 \leq s_{f\rightarrow m} \leq 199.8 \label{L-s_t} \\
    \textrm{Medium}: 143.6 \leq s_{f\rightarrow m} \leq 152.4 \label{M-s_t} \\
    \textrm{High}: 126.1 \leq s_{f\rightarrow m} \leq 133.9 \label{H-s_t} 
\end{align}
It should be noted that, strictly, these values do not differ because the reduced densities of the mRVEs differ, but rather because the characteristic feature length ($\ell$) in the mRVEs differ.
In addition, the unitless values reported here can be scaled to physical units applicable to real foam materials (that are well described by the node-ligament seed geometry) by determining any one of the lengths $s$, $\ell$, or $d$ for a particular sample and scaling the equivalent KRaSTk units.


\subsection{Stochastic model predictions of Duocel properties}
Computed $\mu_E$ and $\sigma_E$ values are intended to be directly analogous to measured properties of metallic foams with structures well modeled by the geometric seed employed here---an assertion that can be directly tested by comparing computed properties to those measured for physical foam samples.
Here we use a commercially available Al metallic foam, Duocel \cite{ERGAerospace:NA:2021}, to demonstrate the predictiveness of the KRaSTk approach.
Duocel, which forms an open foam structure similar to the node-ligament geometric seed used here, is available at 5, 10, 20, and 40 PPI grades, where PPI measures pores per linear inch in two-dimensional cross section.
In the context of the present model, for self-similar structures across Duocel products (that is, under the assumption that all Duocel products express the same connectivity and node-ligament geometries), the average pore diameter (the inverse of PPI) would be proportional to the average length of ligaments bounding the pores.
Therefore, for Duocel materials with self-similar structures, variations in PPI at fixed density are equivalent to variations in average ligament length ($\ell$), and, for fixed physical sample sizes ($s$), to $\lambda$.
Changes in density at fixed PPI effectively represent changes in ligament diameter (strictly, in ligament aspect ratio, as at fixed PPI average ligament length is fixed).

\begin{figure}[h]
    \centering
    \includegraphics[width=1\linewidth]{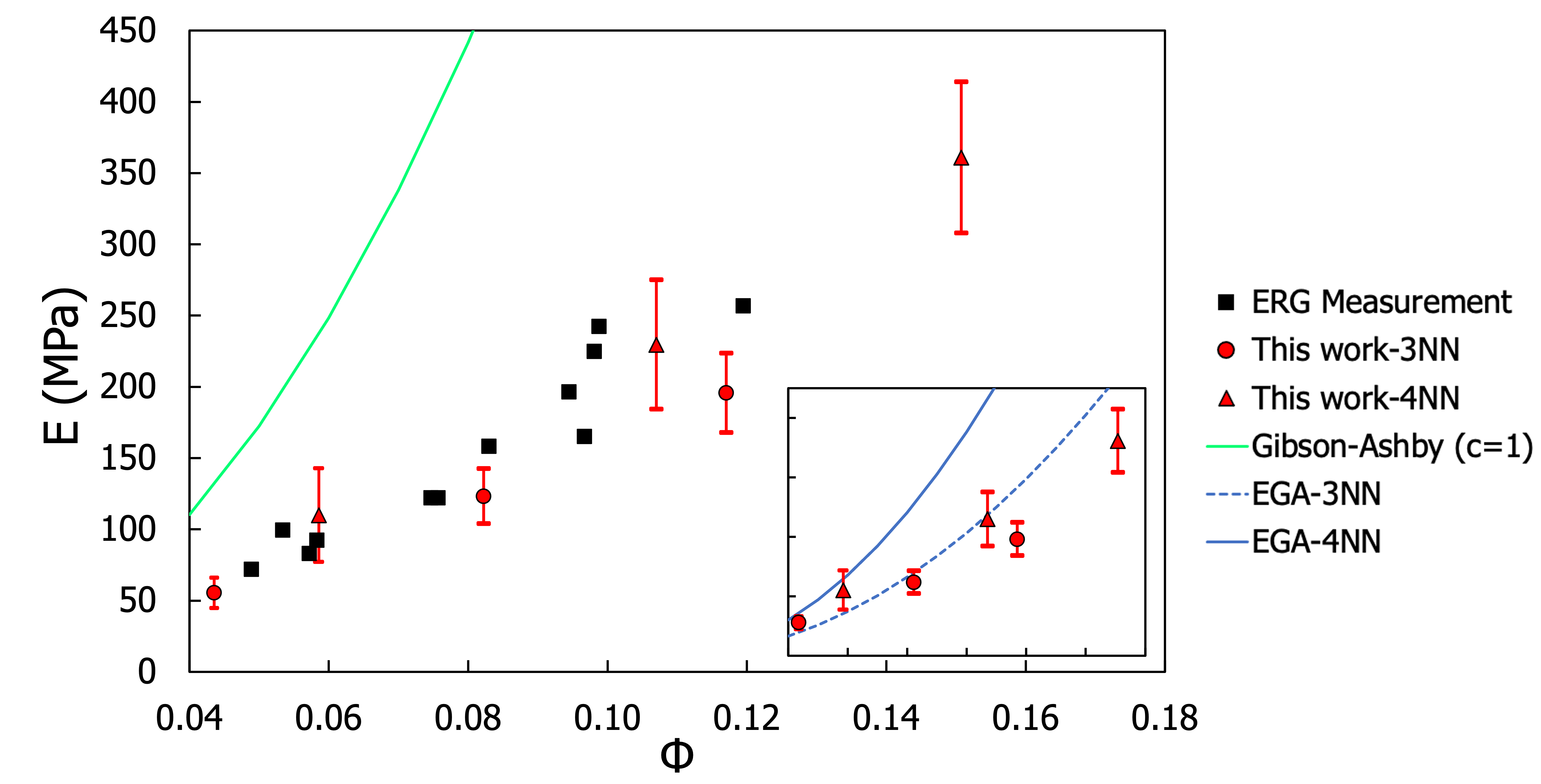}
    \caption{Comparison between KRaSTk prediction and Duocel measurements made by the manufacturer, ERG Aerospace \cite{ERGAerospace:NA:2021}.}
    \label{fig:comparisonwithduocel}
\end{figure}

Fig.~\ref{fig:comparisonwithduocel} plots properties (for both $N_C=3$ and $N_C=4$ mRVEs) computed in this work along with properties of Duocel samples as measured experimentally by the foam manufacturer, ERG Aerospace \cite{ERGAerospace:NA:2021}, according to ISO 13314:2011 \cite{:NA:2011a}.
Samples with a range of densities were tested and each sample had dimensions of approximately 50 mm $\times$ 50 mm $\times$ 65 mm.
Computed properties reported in Fig.~\ref{fig:comparisonwithduocel} are for mRVE sets with $\lambda$ values at or just above the $\lambda_{f\rightarrow m}$ range discussed above--that is, applicable to equivalent physical samples with sizes $\sim10\times$ typical ligament lengths.
Analytic predictions of both the Gibson-Ashby model\cite{Gibson:NA:1997,Ashby:NA:2000} and the Extended Gibson-Ashby expression derived in Ref.~\cite{Seif:AM:2021} for foam stiffness are also shown.

In Fig.~\ref{fig:comparisonwithduocel} red markers indicate $\mu_E$ values and whiskers indicate the standard variability ($[\pm \sigma_E]$).
The inset shows only computed results obtained in this study (red points) and analytic model curves for open-cell foams based on the Extended Gibson-Ashby equation presented in Ref.~\cite{Seif:AM:2021}:
\begin{equation}
    \frac{E^*}{E_s} = 0.49\phi^{2}(N_C^*-3)^{0.41}
\end{equation}
The EGA model was developed to extend the longstanding Gibson-Ashby model for the mechanical properties of porous materials by accounting for the connectivity of node-ligament network structures.
Careful consideration of Fig.~\ref{fig:comparisonwithduocel} reveals that computed elastic properties for mRVEs with $N_C=4$, presented in Table~\ref{tab:connectivty4} in the Appendix, match experimental results both in terms of expected bulk elastic modulus values ($\mu_E$) \cite{ERGAerospace:NA:2021} and in terms of the apparent variability of elastic properties among samples at sizes near $\lambda_{f\rightarrow m}$ ($[\pm \sigma_E]$).
For samples with a reduced density of $\phi\approx0.10$, the scatter or apparent variability in measured properties as reported by ERG is about $\pm 39$ MPa.
At a similar $\phi$, the KRaSTk-computed standard variability is $\pm 45$ MPa.
Similar agreement is observed throughout the considered range of $\phi$ values.
Fig.~\ref{fig:comparisonwithduocel} also shows that computed results for $N_C=3$ fall systematically below measured stiffness values.
While network connectivity is not a parameter directly controlled during Duocel fabrication, 
independent evaluation of network connectivity in Duocel foam by ERG (based on micro-CT volumetric imaging) confirms that Duocel does, in fact, have 
$N_C=4$ \cite{ERGAerospace:NA:2021}.
 
In aggregate, quantitative agreement between computed and measured mechanical properties demonstrates that the KRaSTk approach---taking only the mechanical properties of bulk Al and the node-ligament seed geometry as input---accurately represents the structure of Duocel and quantitatively predicts Duocel mechanical properties.
In turn, this demonstrates that KRaSTk can be applied directly to address the needs of spacecraft and other system designers seeking knowledge of not just the bulk effective properties of Duocel, but also the sample-size-dependent variability in properties and the length scales bounding the feature-scale, mesoscale, and macroscale behavior of Duocel.
As shown here, mean computed values of mRVE mechanical properties quantitatively predict measured Duocel properties, and the standard variability in mRVE properties quantitatively predicts the sample-to-sample variability that should be expected in finite-sized Duocel samples.
Importantly, while the present study does not consider sample sizes large enough to determine the meso-to-macroscale transition (above which no sample-to-sample variability in mechanical properties would be observed), the computed feature to mesoscale transition length ($\lambda_{f\rightarrow m}$) sets a minimum sample size below which Duocel (and similar materials) will no longer exhibit properties (and variability in properties) characteristic of the open-cell foam material itself.
This represents a minimum size threshold for components in any design decisions based on Duocel foam properties, as opposed to properties of individual ligaments and pores.


\subsection{MMOD interaction scale}
The role of $\lambda_{f\rightarrow m}$ as a minimum size threshold for design decisions relating to the use of Duocel can be highlighted by exploring the implications of the feature-to-mesoscale transition on a specific and critical application of Duocel: as the energy dispersing core of open cell metallic foam sandwich panel MMOD shielding.
Given known sizes of MMOD impact volumes, computed $\lambda_{f\rightarrow m}$ values can be used to determine whether MMOD interactions with Duocel can be understood in terms of effective properties of the foam, or whether such interactions necessarily must consider the (microscopic) individual interactions between MMOD particles and individual ligaments and pores in Duocel.
As noted above, while the PPI of Duocel samples should generally scale with average ligament lengths, mRVEs are unitless, characterized by their relative size $\lambda$.
Physical length scales can be assigned to mRVEs at all $\lambda$ values by calibrating to the average ligament length in particular Duocel samples to be considered for use in an application.
Previous studies of Duocel have determined that average ligament lengths in 8\% dense 10-40 PPI samples range from $\sim$1-2 mm \cite{Jang:IJoSaS:2008}.
Using this as a reference for the medium density mRVEs here (which exhibited an average density of 8.1\%), one KRaSTk unit of length is approximately 67-135 $\upmu$m.
Given that the feature-to-mesoscale transition occurs at a common \emph{relative} size scale ($\lambda_{f\rightarrow m} \approx 10$) regardless of reduced density, the minimum sample volume must be about $10\times$ the characteristic ligament length.
This arises directly from the definition of $\lambda$ as the ratio of the sample size to the characteristic feature size.
Since the characteristic ligament lengths are known from previous experiment, the side lengths for each reduced density mRVE can be computed (assuming that one KRaSTk unit is equivalent to 100 $\upmu$m): 19.4 mm (low), 14.9 mm (medium), and 13.0 mm (high).

This materials specific conversion to physical length scales leads to two important results.
First, it provides a justification for our comparison, in the previous subsection, of computed results for mRVEs with sizes (measured via $\lambda$) of close to $\lambda_{f\rightarrow}m$.
Specifically, physical samples of Duocel as tested were 50 mm $\times$ 50 mm $\times$ 65 mm, which is slightly larger than the $s_{f\rightarrow m}$ estimated here.
Moreover, these $s_{f\rightarrow m}$ values have important implications for the use of these materials as components in MMOD shielding.
Revisiting Fig.~\ref{fig:mmodimpacts}, the diameter of a representative MMOD strike is approximately 500 $\upmu$m.
On average this is 30$\times$ \emph{shorter} than the feature-scale to mesoscale transition length for any reduced density foam. 
Therefore, bulk (or even mesoscale) materials properties of Duocel will have limited applicability for determining the response of Duocel foam to MMOD strikes---instead, the individual properties of ligaments and/or pores will dominate the materials response.

\section{Summary and Outlook}
Direct comparison of computed properties of metallic foam mRVEs to experimentally measured properties of commercially available Duocel foam demonstrates that the KRaSTk method with a node-ligament seed geometry allows for quantitatively accurate predictions of both expected average or bulk properties as well as the variability expected to be observed in measured properties as a function of sampling volume.
Leveraging this computational approach, the critical size scale at which foam samples will exhibit characteristic materials behavior ($\lambda_{f\rightarrow m}$, the transition from the feature-scale to the mesoscale) is found to occur at a common sampling volume relative to the size of characteristic structural features (here, the average length of ligaments in an mRVE).
For the node-ligament seed geometry, mesoscale behavior emerges once the side length of the sampling volume approaches or exceeds $\sim10\times$ the characteristic feature size.
Despite a generally applicable \emph{relative} sampling volume, lower density foams transition to the mesoscale at higher absolute sampling volumes.
For Duocel, mesoscale behavior does not emerge until sample volumes exceed $\sim1$ cm$^3$.
This feature-to-mesoscale transition volume sets a minimum size scale for parts or components to exhibit performance representative of the bulk material; moreover, for sample volumes near this transition, intrinsic variability (albeit characteristic, predictable variability) in materials properties will still be observed and must be considered.

For metallic foams generally, computed elastic properties at the low reduced densities considered here ($<$12\%) do not align well with the predictions of analytic theories, including both the longstanding Gibson-Ashby model, which considers only variations in reduced density \cite{Gibson:NA:1997}, and the recently developed Extended Gibson-Ashby (EGA) model, which adds a term accounting for variations in network connectivity \cite{Seif:AM:2021}.
In general, the EGA provides substantially better agreement, though both models could, retroactively, be scaled with arbitrary pre-factors to better coincide with computed (and experimentally measured) results.
Direct KRaSTk calculations, comparatively, taking as inputs only a seed geometry and the bulk elastic properties of the constituent ligament metal, yield quantitatively accurate predictions of both expected or average elastic moduli and the observed variation in measured properties of mesoscale samples.
Additional studies are necessary to assess whether a refined EGA requires a modified form at both low and high densities.

A potential issue related to extending the applicability of the present results to other materials with complex structure centers on determination of the characteristic structural length scale.
For the metallic foam seed geometry used here it is convenient to use the average ligament length, but, for example, in fibrous materials (see e.g., Refs.~\cite{Seif:2ICFVARMEF:2022,Seif:S:2021}) the relevant characteristic length for normalized system size may be the average fiber length or the distance between fiber-to-fiber mechanical contacts.
In addition, for textured materials with orientation dependent structures (e.g., fibrous mats or woven materials) it may be necessary to have a multi-dimensional $\lambda$ accounting for different characteristic lengths in different directions.
Finally, the use of isotropic properties for the constituent material from which structural features are formed (here, polycrystalline Al 6101-T6 for Duocel) should be carefully considered.
If structural feature sizes approach the grain size of a constituent material, care must be taken to properly account for the anisotropic nature of the underlying material.

Regarding the response of Duocel foam (and similar materials) to hypervelocity impacts and MMOD strikes, specifically, the results presented here demonstrate that initial impact volumes are sufficiently small (relative to the key structural features of shielding materials) that the impacting particle does not interact with the bulk or even mesoscale material, but rather with individual elements of the constituent phases---generally voids and ligaments.
Therefore, the materials response of metallic foam shielding material to such impacts cannot be modeled accurately by treating the shielding material as a homogeneous bulk or mesoscale material.

Overall, as noted above, the modeling approach applied here yields quantitatively accurate predictions of both expected bulk homogenized properties as well as distributions of local materials properties as a function of sampling volume (for volumes larger than $\lambda_{f\rightarrow m}$).
Given knowledge of a relevant interaction volume for a process of interest, KRaSTk-computed property distributions represent a direct quantification of both expected properties and uncertainty in materials behavior.
Similar results can be expected for any material with a complex structure that can be represented using a seed geometry appropriate for procedural generation of mRVEs, offering a promising approach for addressing urgent materials modeling needs for component and systems designers seeking to employ such materials in a range of critical applications.

\section*{Acknowledgments}
The research presented here was supported by M.N. Seif's NASA Space Technology Graduate Research Fellowship (Grant Number 80NSSC20K1196). We thank the University of Kentucky Center for Computational Sciences and Information Technology Services Research Computing for their support and use of the Lipscomb Compute Cluster and associated research computing resources.

\newpage
\section*{Appendix}


\subsection*{Calculation of elastic constants}
\noindent This section describes how elastic constants presented here were obtained from FEM-computed strain energies. After first converting strain energies to strain energy densities (i.e., normalized by volume, $s^3$), the stiffness matrix was computed as: 
\begin{align}
    C_{11} &= (2*U_{11})/(\epsilon*\epsilon) \nonumber \\
    C_{22} &= (2*U_{22})/(\epsilon*\epsilon) \nonumber \\
    C_{33} &= (2*U_{33})/(\epsilon*\epsilon) \nonumber \\
    C_{12} &= ((2*U_{12})/(\epsilon*\epsilon))-((2*U_{11})/(\epsilon*\epsilon))-((2*U_{22})/(\epsilon*\epsilon)) \nonumber \\
    C_{13} &= ((2*U_{13})/(\epsilon*\epsilon))-((2*U_{11})/(\epsilon*\epsilon))-((2*U_{33})/(\epsilon*\epsilon)) \nonumber \\
    C_{23} &= ((2*U_{23})/(\epsilon*\epsilon))-((2*U_{22})/(\epsilon*\epsilon))-((2*U_{33})/(\epsilon*\epsilon)) \nonumber \\
    C_{44} &= (2*U_{44})/(\epsilon*\epsilon) \nonumber \\
    C_{55} &= (2*U_{55})/(\epsilon*\epsilon) \nonumber \\
    C_{66} &= (2*U_{66})/(\epsilon*\epsilon) \nonumber 
\end{align}
\noindent Finally, convert between entries in the stiffness matrix (C$_{ij}$) and elastic constants (E, $\nu$, and G).
\begin{align}
    E_1 &= \frac{C_{11}C_{22}C_{33}+2C_{23}C_{12}C_{13}-C_{11}C_{23}C_{23}-C_{22}C_{13}C_{13}-C_{33}C_{12}C_{12}}{C_{22}C_{33}-C_{23}C_{23}}\nonumber\\
    E_2 &= \frac{C_{11}C_{22}C_{33}+2C_{23}C_{12}C_{13}-C_{11}C_{23}C_{23}-C_{22}C_{13}C_{13}-C_{33}C_{12}C_{12}}{C_{11}C_{33}-C_{13}C_{13}}\nonumber\\
    E_3 &= \frac{C_{11}C_{22}C_{33}+2C_{23}C_{12}C_{13}-C_{11}C_{23}C_{23}-C_{22}C_{13}C_{13}-C_{33}C_{12}C_{12}}{C_{11}C_{22}-C_{12}C_{12}}\nonumber\\
    \nu_{12} &= \frac{C_{12}C_{33}-C_{13}C_{23}}{C_{22}C_{33}-C_{23}C_{23}}\nonumber\\
    \nu_{13} &= \frac{C_{22}C_{13}-C_{12}C_{23}}{C_{22}C_{33}-C_{23}C_{23}}\nonumber\\
    \nu_{32} &= \frac{C_{11}C_{23}-C_{12}C_{13}}{C_{11}C_{22}-C_{12}C_{12}}\nonumber\\
    G_{23} &= C_{44}\nonumber\\
    G_{13} &= C_{55}\nonumber\\
    G_{12} &= C_{66}\nonumber
\end{align}

\newpage
\subsection*{Properties of $N_C=4$ mRVEs}
\noindent The distribution parameters for mRVE sets with $N_C=4$ are shown Table~\ref{tab:connectivty4} below.
\begin{table}[hbt]
    \centering
    \caption{The gamma distribution parameters, $a$ and $b$, and corresponding physical quantities, $\mu_E$ and $\sigma_E$, of each mRVE set with $N_C=4$.}
    \begin{tabular}{ccccc}
        \toprule
        & \multicolumn{2}{c}{$\Gamma$ Parameters} & \multicolumn{2}{c}{Elastic Modulus} \\
        \cmidrule(lr){2-3}\cmidrule(lr){4-5}
        Set & $a$ & $b$ & $\mu_E$ & $\sigma_E$ \\
        \midrule
        L/10.3 & 11.19 & 9.82 & 109.89 & 32.85 \\ 
        \midrule
        M/11.6 & 25.60 & 8.97 & 229.63 & 45.39\\
        \midrule
        H/11.1 & 44.80 & 8.04 & 360.22 & 53.82\\
        \bottomrule
    \end{tabular}
    \label{tab:connectivty4}
\end{table}

\clearpage
\bibliography{references}

\end{document}